# All-optical seeding of a light-induced phase transition with correlated disorder

Allan S. Johnson[1,2*+], Ernest Pastor[2,3*+], Sergi Batlle-Porro[2], Hind Benzidi[4], Tetsuo Katayama[5,6], Gilberto A. de la Peña Muñoz[7,8], Viktor Krapivin[7,8], Sunam Kim[9], Núria López[4+], Mariano Trigo[7,8], Simon E. Wall[10+].

1. IMDEA Nanoscience, Calle Faraday 9, 28049 Madrid, Spain

2. ICFO-Institut de Ciencies Fotoniques, The Barcelona Institute of Science and Technology, Av. Carl Friedrich Gauss 3, 08860 Castelldefels (Barcelona), Spain

3. IPR–Institut de Physique de Rennes, CNRS-Centre national de la recherche scientifique, UMR 6251 Université de Rennes, 35000 Rennes, France

4. Institute of Chemical Research of Catalonia (ICIQ-CERCA), The Barcelona Institute of Science and Technology (BIST), Av. Països Catalans, 16, 43007 Tarragona, Spain

5. Japan Synchrotron Radiation Research Institute, 1-1-1 Kouto, Sayo-cho, Sayo-gun, Hyogo 679-5198, Japan.

6. RIKEN SPring-8 Center, 1-1-1 Kouto, Sayo, Hyogo 679-5148, Japan.

7. Stanford PULSE Institute, SLAC National Accelerator Laboratory, Menlo Park, CA 94025, USA.

8. Stanford Institute for Materials and Energy Sciences, SLAC National Accelerator Laboratory, Menlo Park, CA 94025, USA.

9. Pohang Accelerator Laboratory, Jigokro-127-beongil, Nam-gu, Pohang, Gyeongbuk 790-834, Republic of Korea

10. Department of Physics and Astronomy, Aarhus University, Aarhus, Denmark

*These authors contributed equally to this work.

+Corresponding authors: allan.johnson@imdea.org, ernest.pastor@univ-rennes.fr, nlopez@iciq.es, simon.wall@phys.au.dk

**Ultrafast manipulation of vibrational coherence is an emergent route to control the structure of solids. However, this strategy can only induce long-range correlations and cannot modify atomic structure locally, which is required in many technologically-relevant phase transitions. Here, we demonstrate that ultrafast lasers can generate incoherent structural fluctuations which are more efficient for material control than coherent vibrations, extending optical control to a wider range of materials. We observe that local, non-equilibrium lattice distortions generated by a weak laser pulse reduce the energy barrier to switch between insulating and metallic states in vanadium dioxide by 6%. Seeding inhomogeneous structural-fluctuations presents an alternative, more energy efficient, route for controlling materials that may be applicable to all solids, including those used in data and energy storage devices.**

Solid-solid phase transitions are ubiquitous processes in nature that can be induced by changes in temperature, pressure, or applied fields, and are increasingly important for next generation, non-volatile electronics[6–11]. Thermodynamically, phase transitions are often heterogeneous, taking a percolation or nucleation-and-growth pathway[12].

However, phase transitions can also be non-thermally driven with light[4,13]. These light-induced phase transitions have the potential to transform solids in faster and more controlled ways than allowed thermodynamically. In particular, ultrafast light-induced transitions are generally assumed to be homogenous and coherent[4,13]. The long wavelength of the light, relative to the atomic spacing, triggers coordinated and coherent motion of the atoms in many unit cells, driving a coordinated transition in a large region of the sample. The dynamics of the phase transition can be described entirely in microscopic terms and precise optical control over the transition should be possible. Indeed, numerous experiments have shown that light can induce large scale coherent motion in solids[14–17], and recent multi-pulse experiments in quasi-1D and 2D materials have shown that the efficiency of light-induced switching can be modulated by this coherent motion[1,2], bringing the concept of coherent control from chemistry[18–20] to solids.

However, coherent control requires a well-defined long-wavelength and optically-active phonon mode that connects the crystal structures of both phases[1,4]. This excludes many technologically relevant transitions such as those in chalcogenide glasses which exploit the crystalline-to-amorphous transitions for data storage[6,9,10]. It is thus unclear if ultrafast processes will bring any benefit for controlling phase transitions in these materials.

Here we demonstrate an alternative, incoherent, route for material control on the ultrafast timescale. We improve the energy efficiency of the insulator-metal phase transition in $VO_2$ by up to 6% when compared to a single pulse excitation by exciting the sample with two pulses. The first pulse generates both large-scale coherent structural motion and also temporally incoherent, yet spatially correlated, localized lattice displacements. However, we find that the incoherent local modes, rather than the coherent motion, are responsible for lowering the energy barrier for the formation of the metallic phase. Density functional theory suggests that interactions between polarons are responsible for the transient barrier reduction. As this correlated disorder can, in principle, be induced in any solid, the incoherent approach may expand the applicability of ultrafast control to a broad range of materials that cannot be coherently steered.

$VO_2$ is a prototypical system for understanding phase transitions in quantum materials[21]. At high temperatures the system is in a metallic rutile phase (R), but below $T_c \sim 60$ °C the vanadium ions pair and twist around the rutile c-axis resulting in a monoclinic insulating phase (M1). Weak photoexcitation of the M1 phase generates coherent phonons, which dynamically modulate the amplitude of the dimerization and tilting of the vanadium atoms[17,22]. Excitation of the M1 phase above a critical fluence threshold $F_{th}$ drives the system to the metallic rutile phase, R. Initially it was assumed that the coherent motion observed at low fluences continues through the phase transition[23,24], but more recent work showed that the coherent motion is quickly transformed into a broad range of incoherent phonon modes and the transition is driven by disorder[25,26]. This places $VO_2$ between charge density wave systems, which are transformed by a few long-wavelength modes[27,28], and crystalline-amorphous phase transitions in which local uncorrelated distortions drive the transition[8,10,29]. Thus, $VO_2$ is an ideal material in which to test how ultrafast control can be applied in systems showing both coherence and disorder.

**Ultrafast control of a phase transition**

We explore optical control of the phase transition by performing multi-pulse excitation experiments in a high-quality single crystalline sample of $VO_2$. We first use a weak pulse (henceforth labelled *prep* pulse) to *prepare* a non-thermal coherently vibrating state in the M1 phase. Subsequently we use a second pulse (labelled *push*) to further excite the sample after a certain delay, $t_{pp}$, and detect if the phase transition occurred with a third *probe* pulse after a time $t_d$ (**Fig 1a**). **Fig 1b** shows the transient reflectivity as a function of probe delay for a fixed *prep-push* delay, $t_{pp}$ = -700 fs, corresponding to 4 periods of the coherent 5.7 THz mode in $VO_2$. The figure displays the cases where the *push* pulse has both sufficient and insufficient fluence to initiate the phase transition. As the excitation is in-phase, a weak *push* pulse amplifies the coherent motion as we remain in the M1 phase. However, a strong *push* pulse induces a larger change in reflectivity and the coherent phonon is suppressed, indicating a transformation to the R phase.

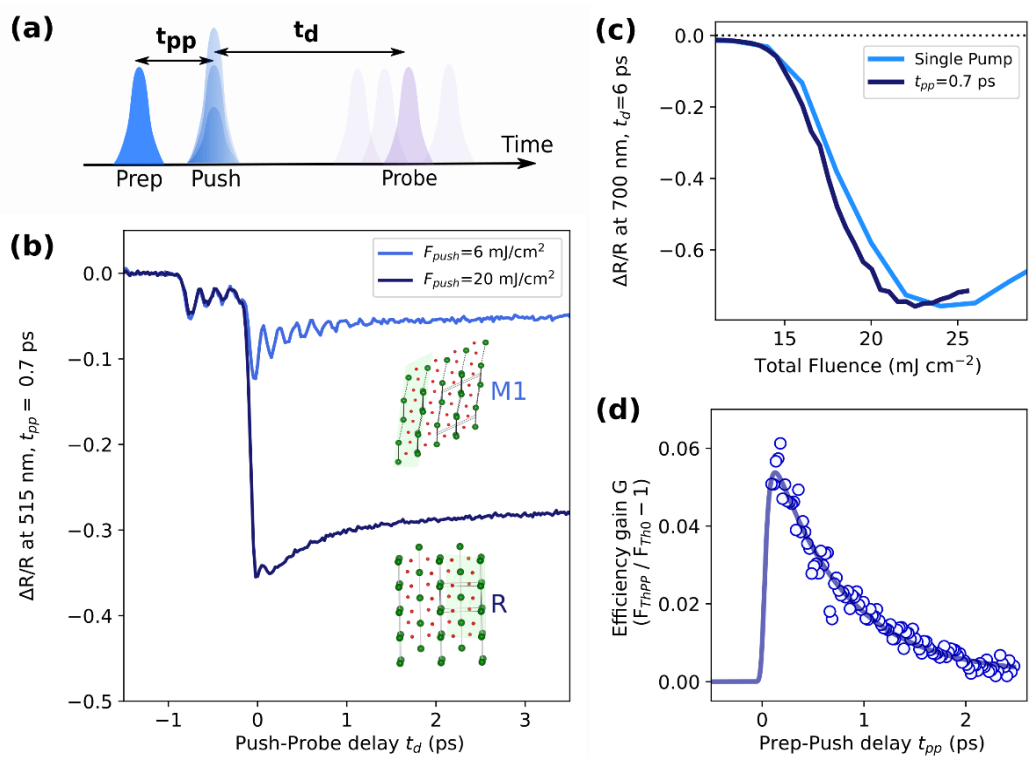

**Fig 1:** (a) Double pump experimental scheme. (b) Transient reflectivity signals as a function of $t_d$. A low fluence *prep* pulse ($F_{prep}$ =5 mJ/cm$^2$) generates coherent phonon oscillations in the M1-phase, while a second excitation can amplify the coherent motion (6mJ/cm$^2$) or drive the system into the R phase (20mJ/cm$^2$). (c) Total fluence dependence of the transient reflectivity for a single pulse and double-excitation measured at $t_d$=6 ps. (d) Relative gain, *G*, in the fluence threshold for a prep-push experiment in comparison to a single pulse experiment, as a function of prep-push delay. There is notable gain at small time delays below the single-pulse fluence requirement. The solid line shows a double exponential fit with 40 fs rise and 700 fs fall time, convolved with the 50-fs experimental resolution. For experimental and fitting details see methods.

**Fig 1c** shows how the transient reflectivity changes as a function of the total incident fluence ($F_{THPP}=F_{prep}+ F_{push}$). For a single pulse excitation, a sudden change in slope at around 15 mJ/cm$^2$ is observed, a marker for the transition's fluence threshold, $F_{TH0}$. Above this fluence, the signal grows rapidly before saturating. Remarkably, for a double-pulse scheme we find that the total energy needed to both initiate and saturate the transition is reduced compared to the single pulse, thus indicating that multiple excitations can lower the energy needed to drive the phase transformation in VO$_2$. We note, the use of a threshold fluence is a different criterion to other works, which instead focused on changes in volume fraction converted[1,2], rather than the energy required to initiate the transition as used in this work.

Next, we explore the dynamics of the threshold reduction. We fix $F_{prep}$ to 5 mJ/cm$^2$, to ensure that a coherent vibrational state is induced, and scan the push fluence, $F_{push}$, for different prep-push delays in order to determine $F_{THPP}(t_{pp})$. In **Fig 1d** we plot the relative efficiency gain, $G=F_{THPP}(t_{pp})/F_{TH0}-1$, as a function of *prep-push* delay, where 0 corresponds to no gain. If coherent motion modulated the threshold energy, we would expect this quantity to oscillate with the phonon period, whereas, if the transition is only limited by the total energy supplied, and not how it is supplied, G should be independent of time. However, we observe neither of these scenarios. Instead, the threshold is strongly reduced at short delays and recovers with an approximately 700 fs time constant before returning to zero after 2 ps, indicating that the threshold energy can be reduced through an incoherent but nonthermal pathway. Although we cannot measure close to time zero due to coherent interference between the *prep* and *push* fields, G must return to zero at $t_{pp} = 0$ to recover the single pulse threshold. Therefore, the gain must rapidly rise within ~125 fs to approximately 6%. We fit this data with a double exponentially with a 40 fs rise time and a 700 fs fall time (see methods).

**Structural dynamics of the photoexcited M1 phase**

Photoinduced changes in the absorption following the initial photoexcitation cannot account for the change in switching efficiency as they are both too small and should display the coherent response in the threshold change (see **SI: Prep-induced change in absorption**). Therefore, we examine if incoherent structural effects could be responsible for the energy saving. **Fig 2a** shows total X-ray scattering measurements, collected at the SACLA XFEL around the (-122) and (-113) M1 Bragg peaks. The M1 peaks measure the long-range order of the vanadium dimers, while the diffuse scattering around the peaks measures momentum dependence of local fluctuations in the structure.

**Fig 2b** shows the long-range structural dynamics after we excite the system with 9 mJ/cm$^2$, just below the fluence required to initiate the phase transition (**SI: Bragg Peak Fluence Dependence)**. The intensity of the Bragg peaks drops rapidly within 65 fs, around our temporal resolution, before partially recovering over a few hundred femtoseconds. On top of these dynamics, the large-amplitude coherent vanadium motion can be observed. The signal decrease can then be ascribed to a combination of coherent displacements and a Debye-Waller-like suppression resulting from incoherent motion[25].

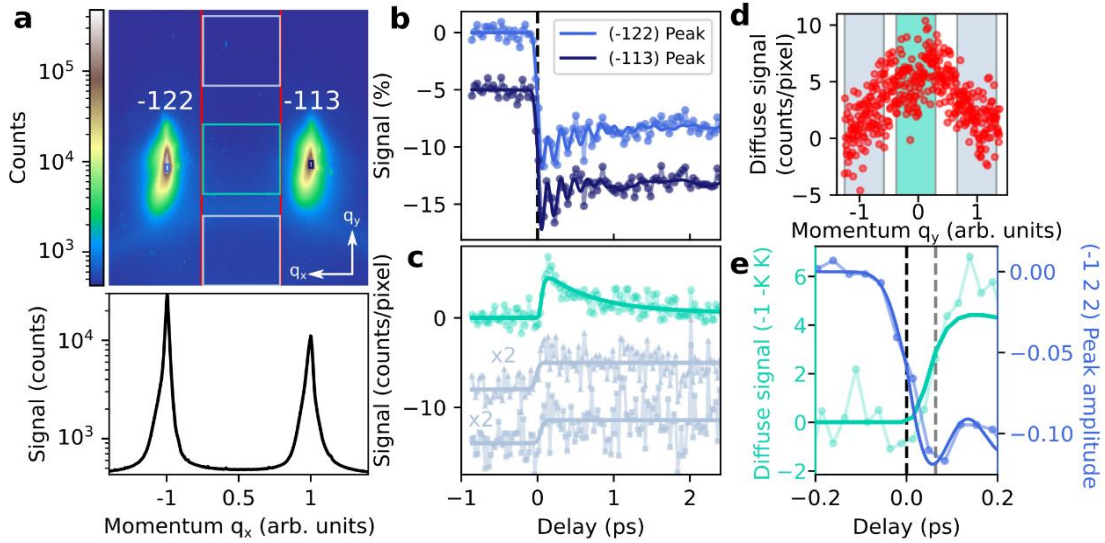

**Fig 2:** (a) Image of 12 keV X-ray scatter from single crystal $VO_2$. Bragg peaks are labeled by the corresponding Miller indices of the M1 phase, while solid boxes correspond to regions of interest (ROIs). Also shown is the signal integrated across the vertical $q_y$ direction. (b) Dynamics near the (-122) and (-113) Bragg peaks below threshold (small blue ROIs, panel a), showing a rapid decrease in amplitude after photoexcitation and strong coherent phonon oscillations. (c) Temporal dynamics of diffuse scatter along the $q_x$=(-1 -K K) direction between the Bragg peaks (green ROI in panel a). Also shown are two regions of interest (light-blue ROIs in panel a) which show only a step-function behavior. (d) Diffuse scatter averaged across the $q_x$ direction at 150 fs delay (red ROI panel a), with the ROIs of the different diffuse ROIs highlighted, showing the transient diffuse scattering is centered between the two peaks at $q_y$=0. (e) Dynamics around time zero for both the normalized Bragg traces and the mean of the diffuse scatter. Dashed lines indicate the half-fall and half-rise times of the two signals, respectively. Solid lines in plots b, c and e are from fits, see methods.

**Fig 2c** analyses the diffuse scattering due to incoherent phonons. We find that the diffuse signal is primarily confined to a stripe of reciprocal space along the (-1 -K K) monoclinic direction spanning wavevectors from (-122) to (-113) (**Fig 2d** and red box in **Fig 2a**). The signal shows a rapid increase upon photoexcitation, followed by a recovery within 2 picoseconds. The dynamics are independent of wavevector along the (-1 -K K) direction within our signal to noise. Outside this stripe-region we observe lower signal levels and a step-like behaviour as shown by the two light-blue ROIs (**Fig 2a**) and traces (**Fig 2c**).

The dynamics of the diffuse stripe-region upon week excitation show a remarkable similarity the gain dynamics shown in Fig. 1(C) in which the system is re-excited after an initial weak excitation. Critically, the increase in the diffuse scattering is delayed by ~50fs relative to the drop in M1 Bragg peak intensity (**Fig 2e**) and we can fit the diffuse dynamics with the same function as the double-pulse gain dynamics in **Fig 1c (See methods).**

## Electronic and atomic structure simulations

Stripe-shaped fluctuations in diffuse scattering result from correlated disorder, where the motion remains correlated in one dimension, but is disordered in the others[30]. In our case the stripes correspond to correlated fluctuations along the body diagonal of the rutile unit cell ($[111]_R$ - rutile notation) as demonstrated in **Supplementary Figure S3**. This suggests that correlated fluctuations along the $[111]_R$ are responsible for the energy saving in the multi-excitation scheme that exhibit the same time dependence (Fig 1(C)). To understand the origins of the correlated fluctuations, we turn to Density Functional Theory (DFT, see methods for details). Within the M1 phase, we find both the ground state (**Fig 3a**) and a meta-stable structurally distorted polaronic state (**Fig 3b**). The polaronic state is primarily electronically confined to two adjacent V sites along monoclinic a-axis (rutile c-axis) with an estimated electronic radius of 3.1 Å, causing a change in V-V bonds (**see SI: DFT Structural Results**). Notably, the polaron exhibits a charge disproportionation where the initial two-$V^{4+}$ charges are separated into one $V^{3+}$ and one $V^{5+}$, stabilized by a second-order Jahn-Teller (SOJT) octahedral distortion that extends beyond the two electronically effected ions.

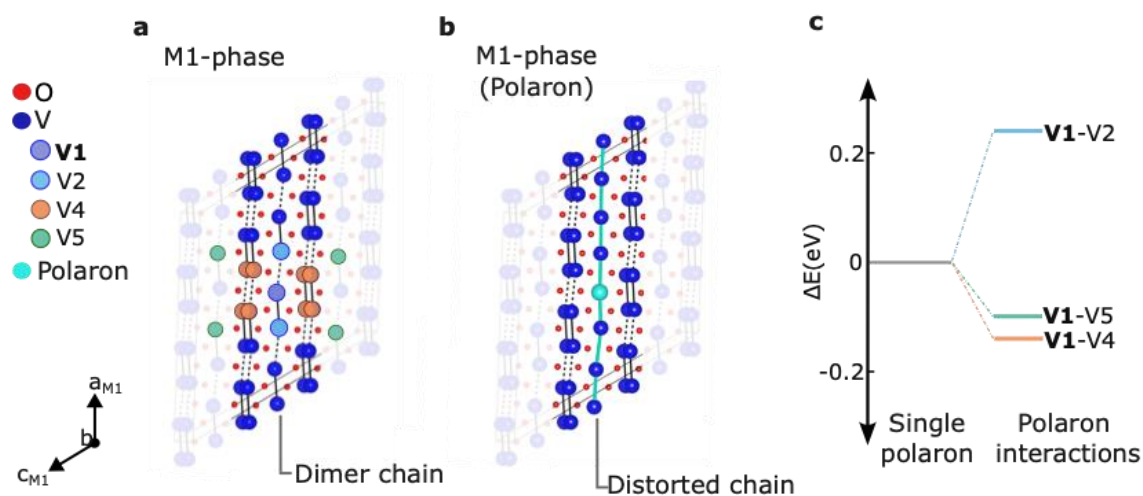

**Fig 3:** (a) Crystal structure of the undistorted M1 phase calculated via DFT and symmetry-equivalent vanadium sites indicated (see methods). (b) Polaron-distorted M1 structure with location of the polaronic distortion. Also highlighted are the relaxed dimer chains resulting from the polaron distortion. (c) Energetics of the polaron-polaron interactions (obtained in a 2x2x2 supercell) for different geometric configurations corresponding to the labelled vanadium atoms in a (see methods).

Such polarons could be expected to form following photoexcitation[31,32], but a local defect alone cannot explain the stripe-shaped diffuse features we measure as these emerge due to correlated fluctuations[30]. Therefore, we examine possible polaronic interactions. For clarity hereafter, we consider the polaron to be located at the $V^{3+}$ site. We find that the energy of the two polaron system strongly depends on the relative location of the distortions (**Fig 3c**). In particular, two polarons at neighboring lattice sites along the monoclinic a-axis (i.e. in the same dimer chain) induce a large energy penalty, whereas two neighboring polarons along the rutile body diagonal (i.e; in neighbouring dimer chains) reduce the energy penalty. This favorable interaction direction, captured by DFT, corresponds to the direction associated with the diffuse stripe features observed **Fig 2**. Based on these observations we propose that cooperative distortions between polarons along $[111]_R$ are responsible for reducing the barrier to the metallic phase.

**Fig 4** summarizes our observations about the control of the phase transition and dynamic modulation of the energy barrier (**Fig 4a**). Photoexcitation first induces coherent motion that is correlated in all three dimensions; this reduces the Bragg intensity of the M1 peaks, without increasing the diffuse scattering, but has no appreciable effect on the threshold energy (**Fig 4b**). Next, after ~50 fs, this fully correlated motion is reduced and only correlations along $[111]_R$ persist giving rise to stripes in the diffuse scatter and which we associate to cooperative polarons (**Fig 4c**). Once formed, the polaronic state lowers the barrier to the metallic R phase. Consequently, driving the phase transition by exciting this state can be achieved with less energy in agreement with our double-pulse gain data. Subsequently, after a few picoseconds, the cooperative polaron interactions are lost and the system recovers a thermal state, which manifests in a more homogenous thermal diffuse pattern. Re-excitation of this state with the *push* beam does not cause any modulation of the transformation barrier (**Fig 4d**).

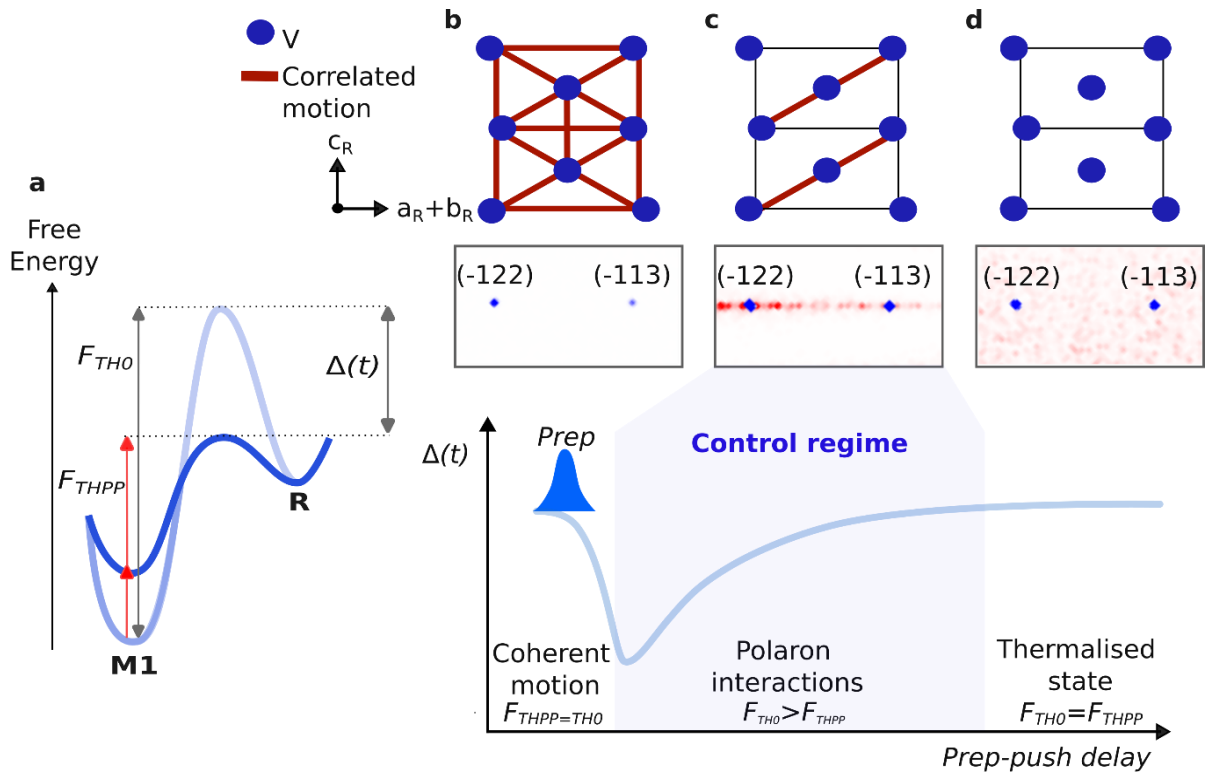

**Fig 4 a** Schematic energy for the M1 and R phase and effect of a multi-pulse excitation. The phase transition from M1 to R is separated by an energy barrier which can be overcome with a single laser pulse that excites the system above a certain fluence threshold ($F_{THO}$). Pre-excitation of the M1 phase with a weak (prep) pulse increases the temperature (free energy), but also creates structural fluctuations which enable a transient change in the energy barrier ($\Delta(t)$) and a reduction of the transition threshold ($F_{THPP}$) in a multi-pulse excitation regime. **b-d** Representation of the atoms intersecting the [111]$_R$ (rutile notation) direction (top) alongside modelled diffuse intensity maps (middle) and the time evolution of the energy barrier (bottom). **b** Initially the phonon displacement is coherent and correlated in three dimensions resulting in a reduction of the M1 Bragg peak intensities without a rise in the diffuse scattering. Coherent motion has no appreciable effect in the energy barrier. **c** After 50 fs the coherence is lost and correlations remain only along the [111]$_R$ direction due to polaronic interactions. This results in the emergence of a stripe-shaped diffuse scatter and a reduction of the energy barrier when the system is re-excited in the strong correlation regime. **d** Finally, the correlations along the [111]$_R$ direction decay as the system thermalizes and the single-pulse energy barrier is recovered.

## Discussion

Recently, theoretical models have suggested that the energy barrier between the insulating and metallic phase in $VO_2$ is reduced if the vanadium ions first untwist before losing the dimerization along the rutile c-axis, rather than moving in a direct and synchronized fashion to the high symmetry state[33]. Our experimental results validate this principle. However, our data reveal that the energy saving results from correlated motion between vanadium ions on *different* c-axis chains. Furthermore, the [111]$_R$ direction is also different from that of the k-vector associated with the phase transition, [101]$_R$, but is consistent with the dominant diffuse scattering observed in metallic phase[34]. This suggests that, for driving the phase transition, it is more important to produce correlated rutile-like distortions along the [111]$_R$ rather than a specific long-wavelength distortions at the [101]$_R$ phase transition wavevector.

While DFT is unable to estimate the lifetime of the polaron state, it seems highly likely that the coherent distortion first breaks down into polarons that are correlated along [111]$_R$ before being converted to random thermal motion. As polaron formation takes ~50 fs (**Fig 2e**), these distortions do not contribute to the phase transition pathway when the system is excited with a single sub-50-fs pulse femtosecond pulse. However, we expect that the phase transition threshold is reduced for longer single pulses as the falling edge of the laser beam could excite polarons created by the leading edge.

Another interesting consequence is that, as the distortion acts to lower the energy barrier between the two phases (**Fig. 4a**), the excess energy needed to drive the transition in the double pump scheme is lower than for a single pump. This excess energy is ultimately converted to heat and thus the resulting metallic phase will be at a correspondingly lower temperature. Further optimization of the energy savings may be possible targeted optimization of how the correlated disorder is generated and harnessed. This might be achieved by more selective excitation of the relevant polaron modes, by tuning the *prep* pulse wavelength or fluence, but further investigation along these lines is needed.

Most notably, correlated disorder can, in principle, be induced in any solid. Consequently, the inhomogeneous seeding strategy we report, might be applicable to a broad range of solids, including those used in energy and data strange applications, for which there is no available optical coherent control route that enables rapid phase changes while still maintaining high energy and thermal efficiency.

## **Methods**

### **Optical Prep-Push Experiments**

A Ti:Sapphire Legend Elite Duo laser supplied by Coherent produces pulses of 35 fs duration at a central wavelength of 800 nm and with an energy of 1 mJ at a 5 kHz repetition rate. The beam is first split into pump and probe pulses, and then the pump into *prep* and *push* using a Michelson interferometer where the power of both arms is controlled independently and a motorized delay stage controls the *prep-push* delay. After this, both *prep* and *push* pass through a chopper and to another delay stage that controls the *push-probe* delay. The *prep* and *push* beams are focused to a spot size 10x bigger than the probe, incident on the sample at an angle of 45 deg. The fluence is calculated using a beam-size extracted by fitting a Gaussian profile to an image recorded by a CMOS camera. The visible probe supercontinuum probe is generated by focusing the 800nm pulses into a sapphire crystal, and subsequently compressed with a pair of chirped mirrors. The reflection of the probe beam from the sample is collected and passed to a spectrometer.

**Total X-ray Scattering Experiments**

X-ray scattering experiments were performed at the SACLA X-ray free electron laser in Japan. A single crystal of $VO_2$ with sample normal along the $[1\ 1\ 0]_R$ (rutile) direction was illuminated by the X-rays at a grazing incidence of ≈0.3° and the resulting scattered X-rays were captured by a multiport charge-coupled device detector placed 106 mm downstream. The $VO_2$ was then photoexcited with optical pump pulses at a central wavelength of 800 nm and 40 fs pulse duration incident at an angle of 5° relative to the X-ray probe. The timing jitter between optical and x-ray pulses was measured and corrected using a spatially-encoded timing tool, resulting in a temporal resolution of 50 fs. Fluence scans were performed to determine the point at which the phase transition was initiated through a bi-linear fit to the drop in Bragg-peak signal level, and subsequent measurements were performed below this fluence (See SI).

**Time-Trace Fitting**

We fit both the efficiency gain $G(t_{pp})$ in double pump experiments and the time dependence of the correlated diffuse scatter along the (-1 –K K) direction by using a double exponential function $S(t) = H(t) * [1 - e^{-t/\tau_1}][Ae^{-t/\tau_2} + B]$, where H is the Heaviside function, $\tau_1$ describes the rise time of the signal while $\tau_2$ describes the decay, A is the amplitude of the transient and B describes the long-term asymptotic change in signal level. The trace was then convolved with a Gaussian of FWHM 50 fs to represent our temporal resolution. We find both the efficiency gain and diffuse signal are well described when $\tau_1$ = 40 fs and $\tau_2$ = 700 fs, as shown in Figures 1d and 2c of the main text. A double exponential format was chosen to best describe the incoherent formation and decay of the structural defects. The dynamics of the Bragg peaks were fit using

$$S(t) = A * \left[\text{erf}\left(\frac{t}{\tau_1}\right) + 0.5\right]\left[Be^{-t/\tau_2} + Ce^{-t/\tau_a}\cos\omega_a t + De^{-t/\tau_b}\cos\omega_b t + E\right].$$

It was necessary to introduce a 15 fs offset in time between the dynamics of the Bragg peaks and that of the diffuse scatter in order to obtain good fitting of the rising edge; this could be due to the need for the long term structural motion to breakdown before the polarons can form, or simply due to the different forms of fitting functions used. This does not, however, affect any of the analysis or conclusions presented here.

**Diffuse scattering simulations**

The reciprocal space maps shown in Fig 4b-d were calculated by performing the Fourier transform of the positions of 20x20x20 vanadium ions. First the diffuse scattering pattern was calculated for the perfect monoclinic structure, which acts as a reference. Then the vanadium ions were moved in three different ways. The coherent simulation moved each V ion by the same distance in a manner that preserves the M1 symmetry. The correlated case moved all ions on the same $[111]_R$ chain by the same displacement, but the displacements were random between chains. The thermal case moved each vanadium ion in a random and independent way. The resulting diffuse patterns were then subtracted from the reference signal.

**Computational methods**

Density-functional theory (DFT) calculations were carried out using the Vienna ab initio simulation package (VASP)[35] and the projector augmented wave (PAW) method[36]. The PBE and HSE functionals were used to model exchange and correlation effects, with HSE06 being the conventional mixing and screening (α = 0.25 and ω = 0.2 Å$^{-1}$) approach. The DFT+U method[37] was used with a U-parameter applied to the V-3d orbitals; We used $U_{eff}$ = U − J = 3 eV with a J value of 1.0 eV. Supercells with (3×3×3) (324 atoms) were used for most calculations, with (6×6×6) and (4×4×4) Monkhorst–Pack grids for PBE, PBE+U and HSE, respectively. K-point meshes were adjusted for larger supercells. A plane wave cut-off energy of 500 eV was used, and the calculations were converged to energies within 10$^{-7}$ eV for electronic structure and ionic relaxations to forces within 10$^{-4}$ eV Å$^{-1}$. In our work, we localized electrons at specific sites in VO$_2$ by distorting VO$_6$ octahedra in the undistorted structure. To explore the stability of the distorted structure, we utilized a combination of GGA+U and the SCAN density functional[38]. We controlled the number of valence electrons and checked the charge density, local lattice environment, and magnetic moment to ensure accurate occupation. We also used chrome pseudopotentials to create wavefunctions with localized electrons. This method is a very efficient approach to form stable polarons. To calculate the polaron formation energy, one needs to compare the total energies of two different electronic states: the delocalized state and the localized state. The delocalized state is typically a band-like state, while the localized state is polaronic. The polaron formation energy ($E_{pol}^{DFT}$) can be defined as the energy difference between the total energy of the delocalized state ($E_{loc}^{undist}$) and the total energy of the localized state ($E_{loc}^{dis}$). Thus, $E_{pol}^{DFT} = E_{loc}^{dis} - E_{deloc}^{undist}$. Determination of the polaron radius by fitting a three-dimensional Gaussian to the electron density to determine the radius where the density decreases to a desired percentage value (see Figure S4). The properties of the polaron like its extension were benchmarked in a large supercell of 3x3x3. However, to calculate the polaron-polaron interactions, a unit cell size of 2x2x2 was used as at least 10 configurations were explored individually. The symmetrical equivalence of different sites (as labelled in Fig3.a) was determined from the M1 unit cell in this reduced cell.

**Data availability**

Source data are available with this paper. Other data that support the findings of this study are available from the corresponding authors upon reasonable request. The DFT computed structures can be found in the ioChem-BD repository:

https://iochem-bd.iciq.es/browse/review-collection/100/64614/b0bbca34e491db60d6fe07b3

**Author contributions**

S. E. W. and M. T. conceived of the project. S. B. P. and E. P. performed the optical measurements and analyzed the data together with A. S. J and S. E. W.

A. S. J., E. P., T. K., G. A. d. l. P. M., V. K., S. K., M.T. and S. E. W. performed diffuse scattering measurements. A. S. J.  and G. A. d. l. P. M. analyzed the X-ray data.

H. B. and N. L performed the polaron calculations.

A. S. J., E. P. and S. E. W. wrote the paper with input from all authors.

**Competing Interests**

The authors declare no competing interests.

**Acknowledgements**

X-ray measurements were performed at BL3 of SACLA with the approval of the Japan Synchrotron Radiation Research Institute (JASRI) (proposal nos. 2018A8007, 2019A8038 and 2019B8075).

ASJ acknowledges support from: ERC AdG NOQIA; MICIN/AEI (PGC2018-0910.13039/501100011033, CEX2019-000910-S/10.13039/501100011033, Plan National FIDEUA PID2019-106901GB-I00, FPI; MICIIN with funding from European Union NextGenerationEU (PRTR-C17.I1): QUANTERA MAQS PCI2019-111828-2); MCIN/AEI/ 10.13039/501100011033 and by the "European Union NextGeneration EU/PRTR" QUANTERA DYNAMITE PCI2022-132919 within the QuantERA II Programme that has received funding from the European Union's Horizon 2020 research and innovation programme under Grant Agreement No 101017733Proyectos de I+D+I "Retos Colaboración" QUSPIN RTC2019-007196-7); Fundació Cellex; Fundació Mir-Puig; Generalitat de Catalunya (European Social Fund FEDER and CERCA program, AGAUR Grant No. 2021 SGR 01452, QuantumCAT \ U16-011424, co-funded by ERDF Operational Program of Catalonia 2014-2020); Barcelona Supercomputing Center MareNostrum (FI-2023-1-0013); EU (PASQuanS2.1, 101113690); EU Horizon 2020 FET-OPEN OPTOlogic (Grant No 899794); EU Horizon Europe Program (Grant Agreement 101080086 — NeQST), National Science Centre, Poland (Symfonia Grant No. 2016/20/W/ST4/00314); ICFO Internal "QuantumGaudi" project; European Union's Horizon 2020 research and innovation program under the Marie-Skłodowska-Curie grant agreement No 101029393 (STREDCH) and No 847648 ("La Caixa" Junior Leaders fellowships ID100010434: LCF/BQ/PI19/11690013, LCF/BQ/PI20/11760031, LCF/BQ/PR20/11770012, LCF/BQ/PR21/11840013).

This project has received funding from the "Presidencia de la Agencia Estatal de Investigación" within the PRE2020-094404 predoctoral fellowship, the Spanish Ministry of Science and Innovation (Ref. No. PID2021-122516OB-I00, Severo Ochoa Center of Excellence CEX2019-000925-S 10.13039/501100011033).

T.K. acknowledges support from JSPS KAKENHI (grant nos. JP19H05782, JP21H04974 and JP21K18944).

E.P acknowledges the support form IJC2018-037384-I funded by MCIN/AEI /10.13039/501100011033.

S.K acknowledges support from National Research Foundation of Korea grant NRF-2019R1A6B2A02100883'

G.A.P.M., V.K., and M.T. were supported by the US Department of Energy, Office of Science, Office of Basic Energy Sciences through the Division of Materials Sciences and Engineering under Contract No. DE-AC02-76SF00515.

## Supplementary Information

### Extraction of the Threshold Fluence

To obtain the fluence threshold we use the large changes in the transient spectrum in the 500-700nm spectral region when the system transforms from monoclinic to insulator. As the spectrum is relatively broad and featureless, often, the change at a single wavelength is used as marker of the transition. While this procedure has been extensively used in the literature since early work in $VO_2$, here we take advantage of broadband detection to avoid having to select a probe wavelength and to incorporate as much experimental data as possible in our assessment of the threshold.

Specifically, we normalize the reflectivity change to that at 550 nm for each delay time. This normalization highlights a change in the spectrum shape and amplitude at fluences above 15 mJcm$^{-2}$, corresponding to the excitation density where the monoclinic modes are lost and large amplitude changes in the un-normalized spectrum occur. Distinctively, at low fluences the signal amplitude above 600 nm is smaller than 1 (i.e lower amplitude than the signal at 550nm in the normalized scale) but it is larger than 1 at high fluences. Next we fit the spectral change between 570-650 nm with a combination of a linear and gaussian function $R = a + b\lambda + A\,e^{-\frac{(\lambda-c)^2}{2s^2}}$, where *a* and *b* are the linear intercept and slope, respectively, and we fix the Gaussian centered at *c* = 610 nm with a standard deviation of *s* = 30 nm. We define the fluence threshold as the fluence at which parameter b changes sign.

We note that analogous results can be obtained by performing the analysis at a single wavelength. This is shown in Figure 1b (main text) displaying a shift in the fluence threshold obtained at one wavelength for a double pump excitation compared to a single pulse excitation. This is also shown Figure S1 displaying the gain in the double pulse experiment obtained using the reflectivity change at 700nm (the data is analogous to that in Figure 1c).

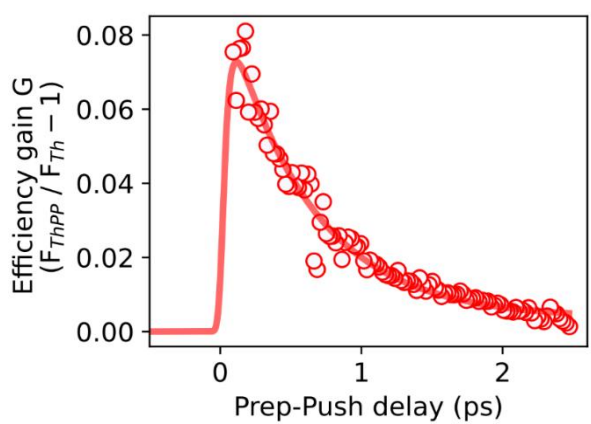

**Figure S1:** Relative efficiency gain for thresholds extracted using a bi-linear fit to the reflectivity data at 700 nm probe wavelength. The overall dynamics are very similar to those obtained using the full spectral data as presented in Figure 1d in the main text.

**Prep-induced change in absorption**

In our measurement we observe a reduction in incidence fluence needed to drive the phase transition. However, this could result from either a reduction in the energy needed to drive the phase transition (our claim), or as more energy is absorbed from the second pulse and thus the total energy supplied is the same.

However, the latter can be easily excluded. First, we note that the coherent and incoherent dynamics are all of a similar magnitude in the optical signal (Fig 1b). Thus, we would expect that the coherent phonon signal would contribute with as strong intensity as the incoherent decay, which is not observed.

Furthermore, we can estimate the pump-induced change of absorption as follows. The summation of the fraction of the energy absorbed ($A$), transmitted ($T$) and reflected ($R$) must equal 1. As our sample is optically thick $T = 0$ so we have

$$R + A = 1$$

Assuming no energy transfer from pump to probe beams (i.e. no stimulated emission) the pump-induced changes mean that

$$R + dR + A + dA = 1$$

$$dA = -dR$$

i.e. the decrease in reflectivity results in an increase in absorption.

The relative change can then be written,

$$\frac{dA}{A} = -\frac{dR}{A} = -\frac{dR}{(1-R)} = -\frac{dR}{R}\frac{1}{R^{-1}-1}$$

At normal incidence $R \sim 0.25$ (ref 25) and the measured $\frac{dR}{R} \sim -0.05$ giving $\frac{dA}{A} \sim 1.6\%$, which is just over a factor of 3 smaller than the observed effect.

**Bragg Peak Fluence Dependence**

Figure S2 shows the fluence dependence of the (-1 2 2) and (-1 1 3) Bragg peaks at a delay of 3 ps. The phase transition is marked by a notable change in the fluence dependence at around 15 mJ/cm$^2$, marked by the deviation from the grey dashed line linear fit to the low fluence data. The below-transition measurements of Figure 2 are taken just below this point at around 9 mJ/cm$^2$ as indicated by the vertical dashed line. Due to penetration depth mismatch between the optical pump and X-ray probe, complete Bragg suppression is not observed. Instead the peak saturates at a residual value of around 40% from the unpumped volume probed by the X-rays; we can relate decreases in the experimental signal to absolute changes in the Bragg peak by removing this offset.

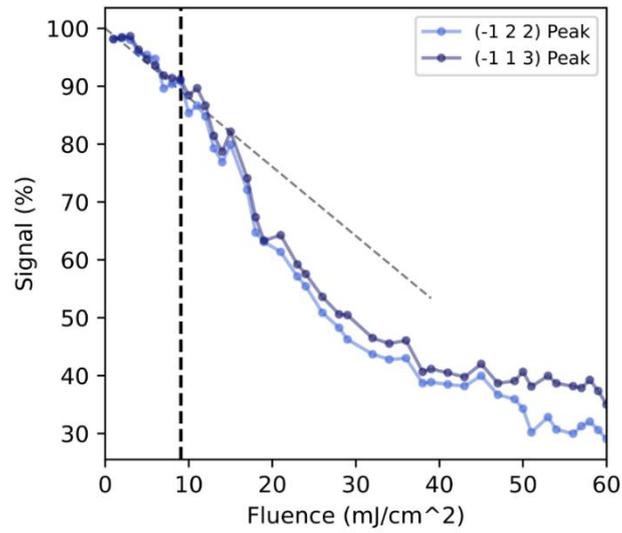

**Figure S2: X-ray fluence dependence of the Bragg peaks.** A low fluence the trend is linear, as shown by the dashed grey fit. Measurement in Figure 2 of the main text were performed at the fluence marked by the black vertical line.

**Correlated disorder**

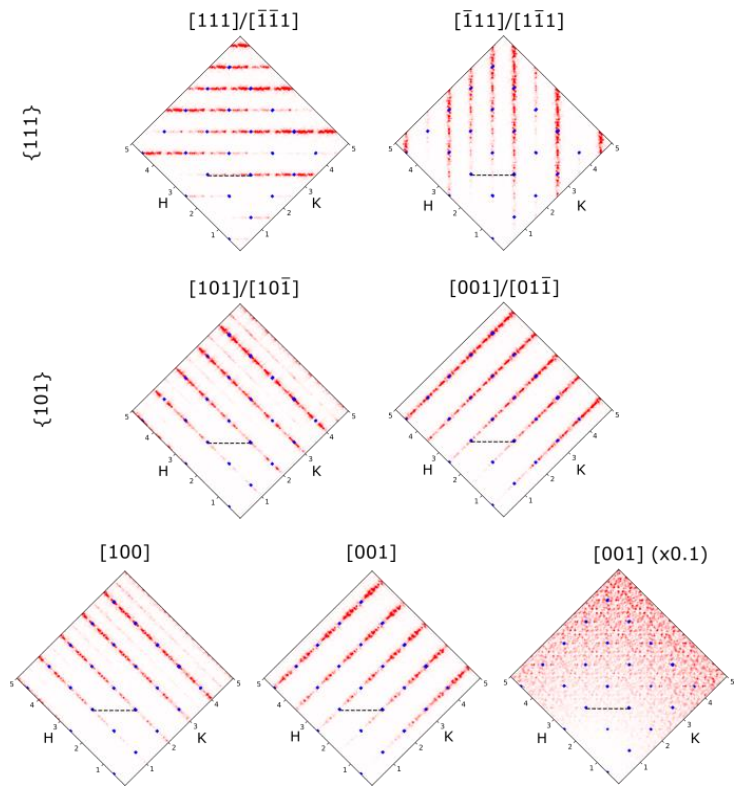

**Fig S3 correlated disorder in the HK-1/2 plane.** Extended reciprocal space maps resulting from correlated disorder along specific directions. Dashed black line corresponds to the stripe direction observed in our experiments. The k-vector, in rutile notation is denoted by the title. The top two rows are grouped by directions that are equivalent in the rutile phase but become inequivalent in the monoclinic phase. It is evident that only [111]/[-1-11] fluctuations match our observations.

**DFT Structural Results**

In the rutile structure, the V atoms create a periodic V chain with a constant V-V bond distance of 2.85 Å. In contrast, in the M1 phase, there are noteworthy discrepancies in the configuration of the V atoms. The V atoms are positioned in a manner that creates dimers alternately and tilts along the c-axis, with a V-V distance of 2.62 Å and 3.16 Å. In our study, we utilized various functionals to optimize the structures of both the R and M1 phases of $VO_2$. We observed that the results obtained from the SCAN meta-GGA, GGA, and reference structural data were relatively similar. To treat the comparison on equal footing, we used the distorted and undistorted structure as initial structure to be performed using the four functionals. Our calculations indicate good agreement between the bandgap obtained from the DFT+U and SCAN calculations of the undistorted structure and reference structural data. In addition to these phases we found the polaronic state in the M1 phase discussed in the main text, whose details are presented in Figure S3.

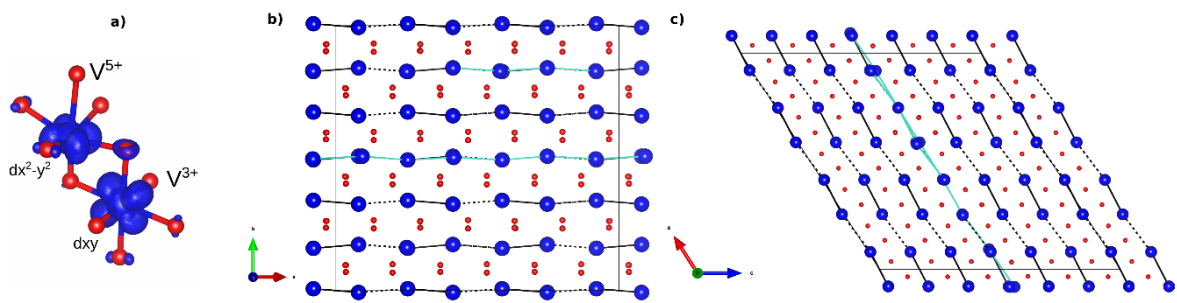

**Fig S4 Visualization of the polaron.** (a) Visualization of polaronic sites and spin density localization at V centers. The labeled 3d orbital indicates significant SOJT distortion in $V^{5+}$ upon polaron formation and weak JT distortion in $V^{3+}$ octahedral. The blue isosurface (blue) is drawn with an isovalue of 0.05 e/Å$^3$. (b) and (c) **ab** and **ac** plan projections of chain rearrangement in V-V dimer highlighting long-range distortion.